\documentclass[]{algotel}
\usepackage[latin1]{inputenc}
\usepackage[francais]{babel}
\author{M. H. Rehmani,\addressmark{1,2}
  A. C. Viana,\addressmark{2}
  H. Khalife,\addressmark{1}
  \and S. Fdida \addressmark{1}}

\usepackage{epsf}\epsfverbosetrue
\usepackage{graphics,epsfig,color}
\usepackage{graphicx,epsfig}
\usepackage{multirow}
\usepackage{slashbox}
\usepackage{alltt}
\usepackage{subfigure}
\usepackage{color}
\usepackage{float}
\usepackage{url}
\usepackage{graphicx}
\usepackage{verbatim} %for giving multiple line comments by mubashir
\usepackage{algorithm}
\usepackage{algpseudocode}
\usepackage{ioa_code}
\usepackage{footnote} % by mubashir
\usepackage{latexsym} % by mubashir
\usepackage{amsmath}
\usepackage{sidecap} %by mubashir
\usepackage{color}
\usepackage{wrapfig}

\title{Adaptive and occupancy-based channel selection for unreliable cognitive radio networks}
\address{\addressmark{1} LIP6/Universit{\'{e}} Pierre et Marie Curie~-- Paris 6, France \\
  \addressmark{2} ASAP/INRIA Saclay - Ilde France sud, France}

\begin{document}
\maketitle

\begin{abstract}\vspace{-0.1cm}
Nous présentons une stratégie dynamique pour le choix de canaux dans un réseau radio cognitif. Notre
stratégie se base essentiellement sur l'estimation de l'occupation du canal radio provoquée par l'activité
des n\oe{}uds primaires plus prioritaires sur les canaux en questions. Notre technique, et grâce à une écoute
permanante sur les canaux, dote les n\oe{}uds cognitifs de la capacité d'inférer les canaux les plus adaptés
à une communication cognitive. En effet, notre stratégie permet d'assurer une fiabilité accrue dans les
communications en évitant les canaux surchargés. Par des simulations, nous validons notre contribution en
montrant qu'elle assure un taux de paquets remis en succès, dans différents scénarios, largement supérieur au
cas où le choix de canaux s'effectue aléatoirement. Par ailleurs, nous montrons également que notre stratégie
permet un nombre supérieur de voisins sur chaque canal que le contexte aléatoire. Cela constitue un résultat
prometteur en faveur de son utilisation dans un contexte multi-saut en présence d'un algorithme de routage. \vspace{-0.5cm}
\end{abstract}
\vspace{-0.45cm}
\section{Introduction}
\label{sec:in}
\vspace{-0.24cm}
Due to limited available spectrum and fixed spectrum assignment policy in today's wireless networks, radio
spectrum is used inefficiently. To counter this issue, Cognitive Radio Networks (CRNs) are designed to allow
cognitive radio (CR) nodes (i) utilizing free parts of unlicensed spectrum as well as (ii) opportunistically
exploiting licensed frequency bands in silent periods of licensed users, called primary radio (PR)
nodes~\cite{survey,mitola}.
%Firstly, CR
%transmissions should not degrade the reception quality of PR nodes and secondly, a CR node should immediately
%interrupt its transmission whenever a neighboring PR activity is detected. Therefore, an unoccupied channel
%is required to allow communication between two or more CR nodes. If a range of channels is available, a CR
%node should select one channel (or more channels) to communicate.

Note that the traffic pattern of primary users is of key importance here : if we know that a particular
channel is heavily used by a PR node and it tends to occupy it for a long period of time, that channel would
be less likely available for a CR node. Because of this PR's high time occupancy, the usability of the
channels by CR nodes becomes uncertain. In addition, since PR nodes operating over different channels have
different activity patterns, depending on the technology they are using and the environment characteristics,
%may have access to different channels,
a hardly predictable and time-variant occupancy can be also perceived. Thus, it is essential and, however,
extremely challenging to correctly select channels allowing reliable communication among CR nodes as well as
channels that increase the number of CR receivers. This constitutes the goal of this paper.

%Since, we are considering infrastructure-less CRNs, there exists several interesting approaches for channel selection~\cite{hai,yang,hoyhtya}, but
%none of these works addressed the key challenge of characterizing channel occupancy incurred by the operating PRs over each
%channel (exploiting historcal-based PR and CR Occupancy for channel selection). 
In this paper, we present a new strategy for {\it adaptive and occupancy-based
channel selection} in CRNs. Our strategy empowers CR nodes with the ability to infer, based on information
regarding PR occupancy, the less occupied channel to use. CR nodes then use the proposed strategy to select
channels for transmission and overhearing. Hence, by exploiting overhearing and PR occupancy properties, our
strategy allows not only increasing the network reliability~-- since less PR-occupied channels will be
selected~-- but also the delivery ratio~-- since a high number of CR nodes will overhear the selected
channel. 
%This is a on-going work and our first try to better understand the channel occupancy in the case of
%CRNs. For this reason, the paper presents the performance evaluation of our approach in a single hop
%scenario. Analyzing and understanding this channel occupancy is crucial to create a basis to build upon in
%order to conceive and develop new approaches for opportunistically route in CRNs.

%Via simulations, we studied the performance of our strategy under a varying number of channels and of PR
%nodes, i.e. occupancy. The simulation results showed that, compared to the case where no PR occupancy is
%considered (i.e. channels are randomly selected to be used), our strategy is practical and effective in
%enhancing the performance of the network in terms of delivery ratio and number of CR receptions.
\vspace{-0.3cm}

\section{Adaptive and occupancy-based channel selection}
\label{sec:proposal}
\vspace{-0.24cm}
%\paragraph{Assumptions:} 
We are considering an ad-hoc
infrastructure-less architecture of Cognitive Radio Network. In this architecture, only CR-CR communications
will occur. Cognitive Radio (CR) nodes are highly frequency agile. Due to PR activity over every channel,
intermittent connectivity of links occurs, requiring CRNs to be highly opportunistic. CR nodes are able to
communicate over multiple channels. 
%We assume that external mechanisms will provide to CR nodes the required
%information from the physical and/or MAC layers to detect PR nodes presence, which consequently, will be used
%to compute occupancy. 
Within the network, two nodes can communicate if they use at least one common channel
and if they are within the transmission range of each other. The spectrum used by PR and CR nodes consists of
N Channels $C=\{1,2,3,\ldots,N\}$, where channels availability for CR nodes is highly time variant and
dependent on the PR activity variability over every channel. \vspace{-0.3cm}

%\paragraph{Our strategy description:}
After sensing the channels in the CRN, each CR node makes a list of detected channels in the network. Each CR
node then classifies the sensed channels in a decreasing order of availability, and selects for transmission
and/or overhearing the best weighted channel. Channels' weight $P_{channel}$ is calculated based on PR \emph{and} CR nodes'
occupancy : \vspace{-0.25cm}
\[
\label{pequation} {P_{channel} = e^{-PROccupancy} \times CROccupancy } \vspace{-0.19cm}
\]
%$PROccupancy$ can be calculated based on the PR nodes' state (i.e. if they are active~-- ON~-- or not~-- OFF). The total number of PRs is evenly
%distributed among existing channels. PR nodes over every channel switch between ON/OFF states with a
%probability of 0.5. The $PROccupancy$ per channel is then calculated based on the number of active PRs at
%each CR transmission,
%While CROccupancy is given by: \vspace{-0.15cm}
\[
\label{pequation1} {CROccupancy = 1 - PROccupancy } \vspace{-0.15cm}
\]
Since PR and CR occupancy are inversely proportional, selection of the less occupied
channel by PR nodes also implies the selection of the most used channel by CR nodes. Hence, the use of the
best channel given by the highest value of $P_{channel}$ allows improving network reliability as well as
increasing the delivery ratio.  \vspace{-0.6cm}
% The following section evaluates the performance of our strategy in terms of
%these two metrics.

\section{Simulation Results and Conclusion}
\label{sec:simulation}
\vspace{-0.3cm}
%\paragraph{Simulation design and metrics:} %We implement our proposal in a simulator written in C++. 
To assess the
performance of the proposed approach, %we varied the number of channels and of PR nodes in the network. 
two performance metrics are evaluated: (i) the average delivery ratio, which is ratio of packet received by a
particular CR node over total packets sent in the network and (ii) the average number of receivers, which is
the real quantity of receivers per node id. In order to simulate message losses, a probability of receiving a
message is assigned to each channel. Results
shown below are generated for an average of 500 simulations, along with 95\% of confidence interval, for a network of 10 CR nodes.% Two
\vspace{-0.4cm}

\begin{figure}[htbp]
    \begin{center}
    \subfigure[]
    { \label{fig_sim_3}
        \includegraphics[width=6.3cm]{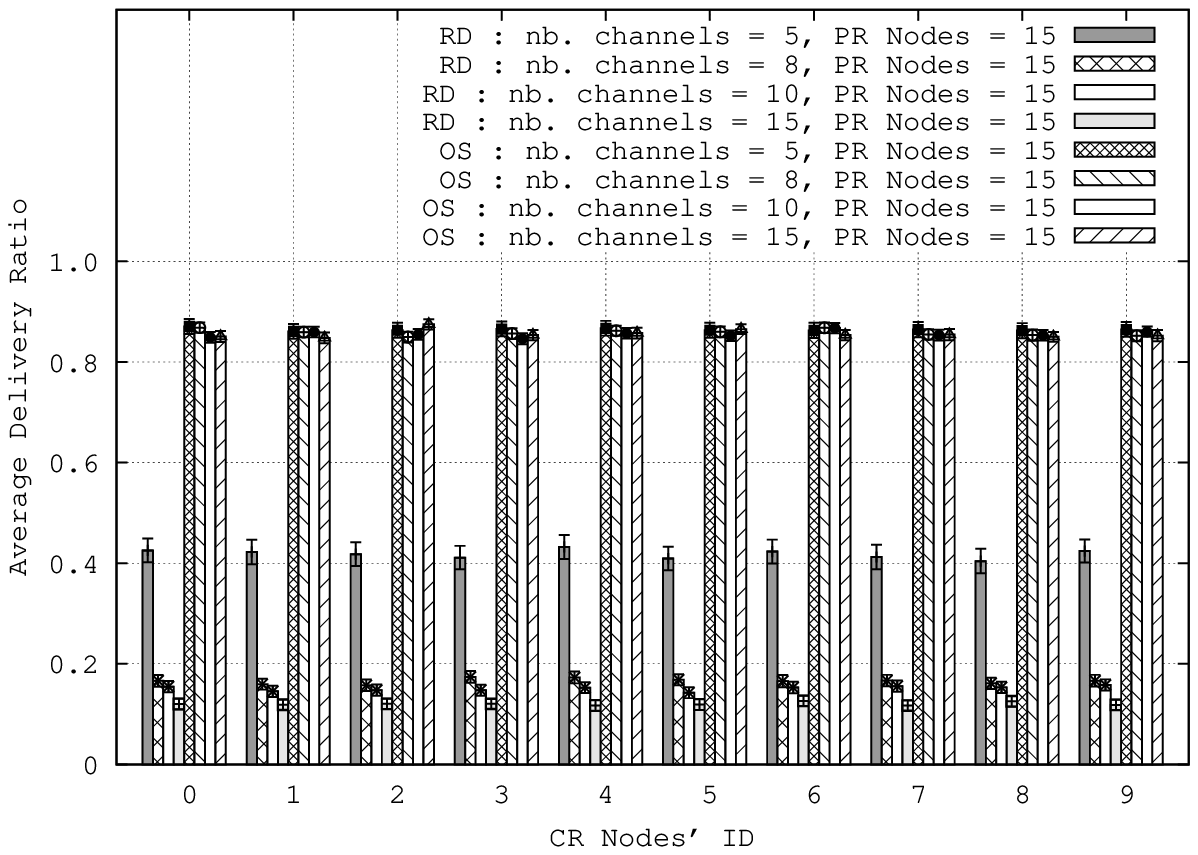}
    }
    \subfigure[]
    { \label{fig_sim_4}
        \includegraphics[width=6.3cm]{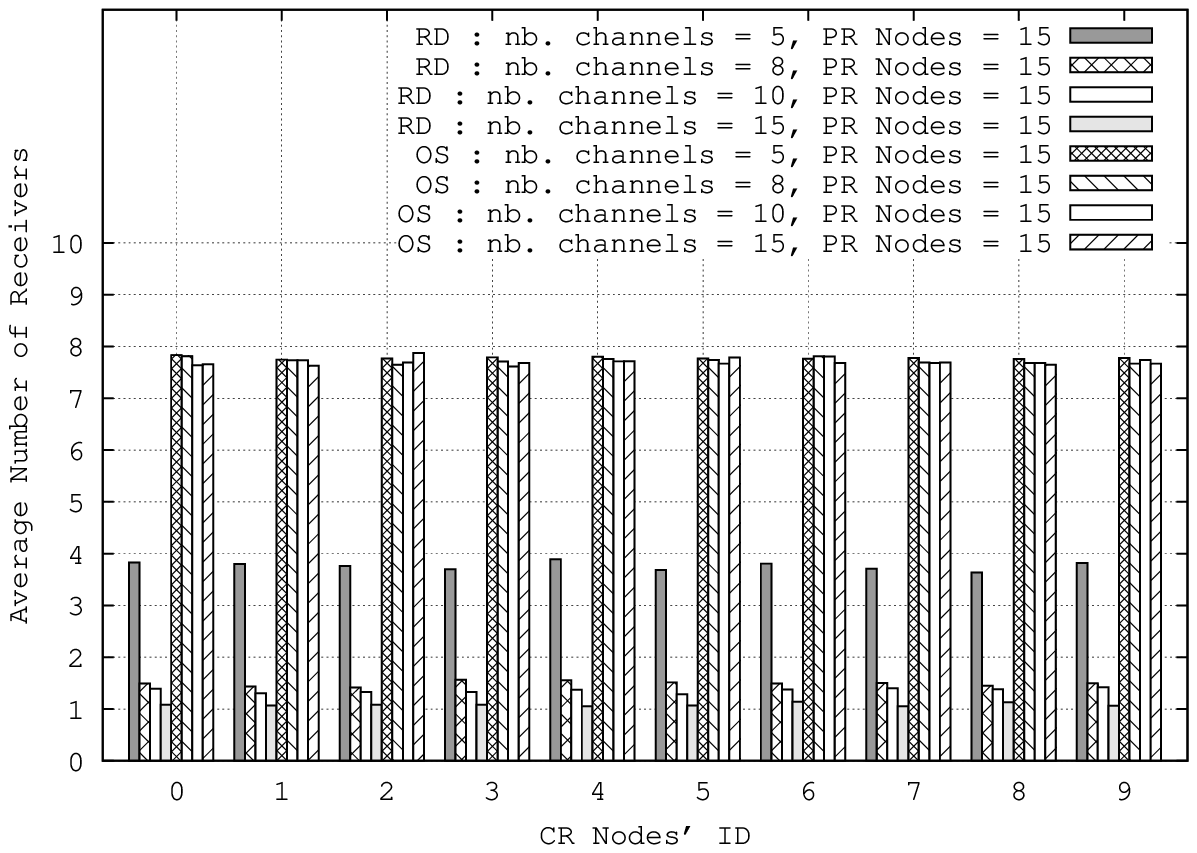}
    }
%    \subfigure[]
%    { \label{fig_sim_5}
%        \includegraphics[width=3cm]{./Figures/Graphs_17_v2/CI_17_Feb_2.eps}
%    }
%    \subfigure[]
%    { \label{fig_sim_6}
%        \includegraphics[width=3cm]{./Figures/Graphs_17_v2/Receivers_17_Feb_2.eps}
%    }
 %   \caption{(a)(c)(e) Graphs showing the average delivery ratio as a function of CR nodes' IDs. (b)(d)(f) Graphs showing the number of receivers as a function of CR nodes' IDs}
%\vspace{-0.4cm}\caption{Average delivery ratio and number of receivers in a CRN with 10 CR nodes, when using
%the metric {\it without prior information about channel condition}, for random (RD) and our adaptive strategy
%(OS).} \label{fig_without_prior_info}
\vspace{-0.4cm}\caption{Average delivery ratio and number of receivers in a CRN, for random (RD) and our adaptive strategy
(OS).} \label{fig_without_prior_info}
\end{center}
\end{figure}

\vspace{-0.6cm}

Fig.~\ref{fig_without_prior_info} compare random (RD) and our strategy (OS) results for the delivery ratio and number of receivers of these two strategies, as a function of the CR nodes' ID. The results
show how the number of channels impacts the average delivery ratio and the average number of receivers for
the two strategies. In particular, by selecting the channel with highest $P_{channel}$, CR nodes are trying
to access best channel in terms of occupancy, for transmitting and overhearing. Hence, as shown in
Fig.~\ref{fig_without_prior_info}, our strategy considerably increases the delivery ratio and the number of
receivers. Our strategy guarantees the delivery of 80\% to 85\% of messages,
contrarily to less than 45\% to 40\% for the random strategy (cf. Fig.~\ref{fig_sim_3}). Note that the increase of the number of channels has a
negative impact on the delivery ratio for the random strategy. This is due to the fact that the probability
of a CR node to randomly select for overhearing, the same channel that another CR node has randomly selected
for transmitting, decreases with the increase of the number of channels. Moreover, our strategy guarantees
from 38\% to 40\% more nodes reception than with the random strategy (cf. Fig.~\ref{fig_sim_4}). Our solution increases the delivery ratio and the number of receivers since it selects the channel with low primary activity and have a high probability to encounter overhearing CR nodes.

%As the number of channels increases, average
%delivery ratio decreases in the case of random strategy and considerably increases in our proposed strategy
%(cf. Fig.~\ref{fig_sim_3}). In addition, our strategy also shows a superior performance in terms of number of
%receivers when compared to the random strategy (cf. Fig.~\ref{fig_sim_4}). 

In summary, our experiments reveal that, by increasing the number of channels, our proposed strategy
considerably increases the average delivery ratio and the average number of receivers, when compared to the
random strategy. In addition, contrarily to random strategy, our proposed strategy guarantees obtaining high
delivery ratio proportional to the number of available channels. This shows how good our strategy is in
allowing CR nodes taking profit of available channels.

\vspace{-0.5cm}
%\section{Conclusion and Future Work}
%\label{sec:conclusion}
%
%The traffic pattern and activity variability of PR nodes in CRNs highly impacts the communication reliability
%among CR nodes. Therefore, this paper presents a strategy for adaptive and occupancy-based channel selection
%for transmission and overhearing of CR nodes. With our strategy and by exploiting occupancy properties, CR
%nodes are able to use less occupied channels with a high number of overhearing CR nodes. This allows
%increasing both network reliability and delivery ratio. 
%In a single-hop scenario, our simulation results
%showed that our strategy significantly enhances the performance in terms of delivery ratio and number of CR
%receptions, when compared with random channel selection strategy. This results are encouraging and represents
%our first investigations for the design of opportunistic forwarding protocols. Therefore, as future works, we
%intend to consider multi-hop scenarios and opportunistically forwarding mechanisms allowing channels
%selection also based on good next-hop CR detection.

%\vspace{-0.2cm}

\nocite{*} \footnotesize{
\bibliographystyle{alpha}
\bibliography{CRmubashir}
\label{sec:biblio} }

\end{document}